\documentclass[9pt,twocolumn,twoside]{opticajnl}
\journal{opticajournal} 

\setboolean{shortarticle}{true}

\usepackage{lineno}
\usepackage[utf8]{inputenc}
\usepackage{amsmath,amssymb}
\usepackage{booktabs}
\usepackage{tabularx}
\usepackage{hyperref}
\usepackage{enumitem}

\title{Urban Mobility and Demand Dynamics: A Telemetry and Ticketing Baseline for Niterói's Bus Network}

\author{Evaldo Alves de Brito Júnior}
\author{Luiz Sergio Pompeu Alves Filho}
\author{Claudio E. C. Campelo}

\affil{Systems and Computing Department, Federal University of Campina Grande (UFCG), Campina Grande, PB, Brazil}

\setboolean{displaycopyright}{false} 

\usepackage{enumitem}
\setlist[enumerate]{topsep=2pt,itemsep=0pt,parsep=0pt,partopsep=0pt}

\begin{document}

\maketitle

\section{Introduction and Motivation}

The public transportation networks of rapidly expanding urban centers confronts a complex task in balancing operational efficiency with dynamic commuter needs, which in face of it data-driven approaches providing multi-perspective datasets offer critical insights into the city's systemic demand.

In this work, we utilize the NetMob 2026 Data Challenge dataset for the city of Niterói in Brazil ~\cite{domingos2026netmob}, to investigate the structural dynamics of the municipal bus network. Specifically, our research framework is driven by three interconnected urban mobility objectives:

\begin{enumerate}
    \item \textbf{Urban Movement Patterns:} To determine whether passenger boarding logs trace periodic, routine, and predictable movement patterns across distinct urban zones in Niterói.
    \item \textbf{Operational Anomalies:} To analyze vehicle trajectory profiles via GPS telemetry and detect both scheduled and unscheduled stops, besides possible bottlenecks.
    \item \textbf{Predictive Modeling:} To develop and evaluate a supervised learning framework capable of predicting continuous bus trip durations.
\end{enumerate}

\section{Dataset and Preprocessing}
Our analytical framework relies on the parsing and synchronization of two primary data repositories provided by the NetMob 2026 Data Challenge: the mobility telemetry streams and the smart card electronic ticketing registers. 

\subsection{Real-Time Bus Tracking}

The mobility dataset comprises real-time GPS telemetry collected from public buses operating in Niterói, Rio de Janeiro, Brazil. The data was obtained through the Mobnit API and follow the General Transit Feed Specification (GTFS), extended with real-time vehicle-position attributes. All recorded bus trips were stored in daily CSV files covering the period from March 11 to March 31, 2026. Each row represents a single vehicle observation, recorded at an approximate frequency of one observation every 15–30 seconds per vehicle. The dataset contains the attributes detailed in Table~\ref{tab:bus_tracking}.

\begin{table}[htbp]
    \centering
    \caption{Attributes of the Real-Time Bus Tracking Dataset}
    \label{tab:bus_tracking}
    \begin{tabularx}{\linewidth}{lX}
        \toprule
        \textbf{Attribute} & \textbf{Description} \\
        \midrule
        \texttt{id} & Unique vehicle identifier \\
        \texttt{timestamp} & Observation timestamp in BRT (UTC$-3$) \\
        \texttt{tripId} & GTFS trip identifier \\
        \texttt{lat} & Vehicle latitude in WGS84 decimal degrees \\
        \texttt{lng} & Vehicle longitude in WGS84 decimal degrees \\
        \texttt{heading} & Vehicle heading in degrees, measured clockwise from North \\
        \texttt{lineId} & Public-facing route identifier (e.g., 24, 49.1) \\
        \texttt{lineName} & Full route name, representing the origin–destination pair \\
        \texttt{headsign} & Destination displayed on the front of the vehicle \\
        \texttt{direction} & Trip direction: 0 (outbound/Ida) or 1 (inbound/Volta) \\
        \bottomrule
    \end{tabularx}
\end{table}

Several gaps and irregularities were identified in the telemetry data. On March 18 and 19, 2026, the telemetry system was unavailable, and therefore no observations were recorded for these dates. March 20 presents a reduced volume of observations, immediately following the outage. On March 22, the temporal interval between consecutive observations was occasionally greater than the expected 15-second sampling interval. Finally, March 28 contains fewer observations than expected.

Overall, the dataset comprises 527 unique vehicles, 7,675 unique \texttt{tripId} values, 72 unique \texttt{lineId} values, and 94 unique \texttt{headsign} values.

As part of the preprocessing procedure, the raw timestamp strings were converted into standardized datetime objects. This transformation established a consistent temporal representation and enabled the extraction of temporal features such as calendar date, day of the week, operational period, and observation time at the required temporal resolution.

\subsection{Passenger Boarding Transactions}
The ticket dataset comprises passenger transaction records from the public bus system in Niterói, Rio de Janeiro, Brazil. Each record represents a single passenger boarding event. The data were stored in daily CSV files covering the entire month of March 2026. The dataset contains the attributes detailed in Table~\ref{tab:ticket_data}.

\begin{table}[htbp]
\centering
\caption{Attributes of the Passenger Boarding Transactions Dataset}
\label{tab:ticket_data}
\begin{tabularx}{\linewidth}{lX}
\toprule
\textbf{Attribute} & \textbf{Description} \\
\midrule
\texttt{transaction\_date} & Boarding timestamp in BRT (UTC$-3$) \\
\texttt{anon\_user\_id} & Anonymized passenger identifier \\
\texttt{fare\_type} & Payment modality \\
\texttt{vehicle\_number} & Unique vehicle identifier \\
\texttt{company\_number} & Bus operator identifier \\
\texttt{route\_detail\_id} & Internal identifier for the route variant \\
\texttt{integration\_flag} & Indicates whether the boarding was part of a transfer \\
\texttt{card\_type} & Passenger category \\
\texttt{debited\_amount} & Fare amount charged, in Brazilian reais (BRL) \\
\texttt{view\_type} & Full route name \\
\texttt{route\_name} & Short route identifier (e.g., 44) \\
\texttt{day\_type} & Day classification: weekday, weekend, or holiday \\
\texttt{day\_period} & Time-of-day category (Dawn, Morning, etc.) \\
\bottomrule
\end{tabularx}
\end{table}

Overall, the dataset comprises approximately 6.7 million passenger transactions, associated with 556 unique vehicles and 58 unique \texttt{route\_name} values. The \texttt{day\_period} attribute divides the day into four discrete operational periods: 00:00–05:59, 06:00–11:59, 12:00–17:59, and 18:00–23:59.

As part of the preprocessing procedure, the \texttt{transaction\_date} field was converted from its raw representation into standardized datetime objects. This transformation provided a consistent temporal representation of passenger transactions and enabled the extraction and analysis of temporal characteristics, including calendar date, day of the week, time of day, and operational period.

\section{Methodology}

\subsection{Spatiotemporal Demand Profiling and Geometry Mapping}
To characterize routine commuter patterns (Objective 1), we used the Passenger Boarding Transactions dataset as the primary source of demand information and analyzed passenger volumes across route lines and operational day periods. The Bus Tracking dataset was subsequently used to provide the spatial representation of the corresponding lines. Since the two datasets use different identifiers for routes, records were matched using \texttt{lineId} in the vehicle tracking data and \texttt{route\_name} in the ticket dataset. Therefore, the analysis was restricted to lines present in both datasets, resulting in an intersection of 48 validated routes.

Then, to spatially represent passenger demand along the physical transit network, the aggregated number of transactions for each five-hour period was associated with the corresponding geographic coordinates of the route obtained from the vehicle tracking data. This procedure assigns the cumulative demand observed during each period to the spatial extent of the corresponding route, allowing the resulting demand patterns to be represented geographically.

Also, to reduce the effects of minor GPS positional variations in the visualization, vehicle coordinates were rounded to a fixed number of decimal places before spatial aggregation. This procedure establishes a regular spatial grid with an approximate spatial resolution of 11~m. Coordinates for which no corresponding passenger transaction records were available were subsequently excluded before generating the network-based visualizations.

\subsection{Operational Stop Validation and Spatial Anomalies}
To detect and analyze both scheduled and unscheduled stops (Objective 2), we developed an algorithm to identify recurring vehicle dwell events across Niterói’s public transport network. A dwell event was defined when the spatial displacement between consecutive observations of the same vehicle remained below a threshold of $10^{-5}$ decimal degrees (approximately 1.1~m) for at least three consecutive GPS observations, corresponding to an average temporal interval of $\Delta t \ge 30$~s.

Relaxing the criterion to only two consecutive observations would substantially increase the number of detected events, accounting for approximately 76\% of all detected stops. However, this configuration was not adopted because of the higher likelihood of false detections caused by GPS positioning errors.

Under the adopted spatial-temporal constraints, the resulting maximum displacement over the observation interval corresponds to an approximate velocity threshold of $v \le 0.04$~m/s (approximately 0.13~km/h), thereby filtering out short movements and minor positional fluctuations while retaining sustained vehicle dwell events.

\begin{figure}[htbp]
\centering
\includegraphics[width=\linewidth]{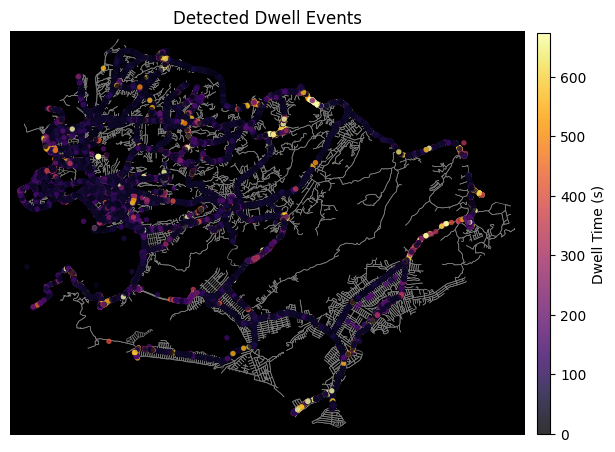}
\caption{Spatial distribution and duration (seconds) of algorithmically detected public transit vehicle dwell events across Niterói.}
\label{fig:dwell_events}
\end{figure}

Using this criterion, a total of 961,281 dwell events were detected. As illustrated in Fig.~\ref{fig:dwell_events}, the duration of the detected events exhibits substantial variability, with a standard deviation of 157.04~s, a mean of 102.42~s, and a median of 46~s. The substantial difference between the mean and median indicates a right-skewed distribution, with a relatively small number of long-duration dwell events contributing disproportionately to the mean.

\begin{figure}[htbp]
\centering
\includegraphics[width=\linewidth]{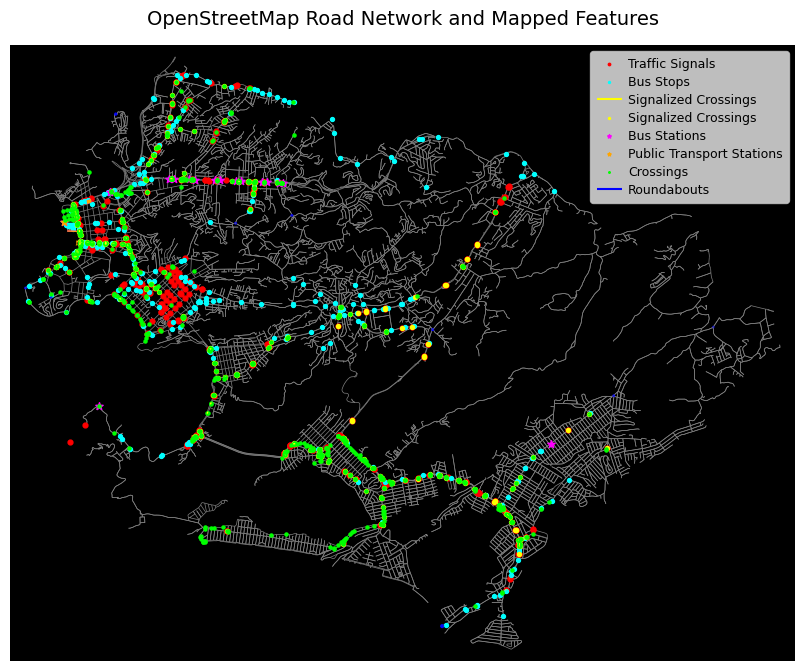}
\caption{Distribution of OpenStreetMap's (OSM) transit infrastructure categories.}
\label{fig:osm_features}
\end{figure}

To investigate the spatial context of these events, the detected dwell locations were compared with relevant OpenStreetMap (OSM) features located along the bus routes in Niterói (Fig.~\ref{fig:osm_features}). The analyzed infrastructure categories included traffic signals, bus stops, signalized crossings, bus stations, public transport stations, crossings, and roundabouts. This analysis was used to assess the spatial association between detected dwell events and transportation-related infrastructure, providing a basis for distinguishing locations potentially associated with scheduled stops from those occurring in other traffic or operational contexts.

Finally, the temporal consistency and distribution of dwell events were analyzed across different days of the week. This analysis was conducted to identify recurring patterns in dwell frequency and duration and to assess whether the detected events exhibit systematic variations according to the day of operation.

\subsection{Continuous Trip Segmentation and Predictive Modeling}
To construct the predictive framework (Objective 3), the telemetry streams were processed sequentially to partition continuous vehicle trajectories into distinct trip segments. Initially, the dataset's \texttt{tripId} was considered as a potential unique identifier for individual trips. However, inspection of the data showed that \texttt{tripId} represents a persistent route itinerary rather than a single operational run. In particular, the same bus may return to a previously observed \texttt{tripId} after operating under a different identifier, indicating that \texttt{tripId} alone cannot reliably delimit individual trips.

Therefore, an empirical analysis was conducted to determine which available vehicle attribute provided the most suitable criterion for defining individual trip segments. Four features were evaluated: \texttt{lineName}, \texttt{tripId}, \texttt{direction}, and \texttt{headsign}. For each feature, a new trip segment was initiated whenever its value changed for the same vehicle. The resulting segments were then compared in terms of their number and duration statistics. The results showed substantial differences among the four segmentation criteria. Using \texttt{lineName} produced only 4,217 trip segments and an average duration of 2,419.91~min, indicating that this attribute changes too infrequently to represent individual trips. In contrast, using \texttt{tripId}, \texttt{direction}, and \texttt{headsign} resulted in 46,148, 69,315, and 72,086 segments, respectively. Among these alternatives, \texttt{headsign} produced both the largest number of segments and the lowest mean (96.11~min) and median duration (45.83~min), suggesting a finer-grained and more plausible representation of individual vehicle runs. Based on these results, a new trip segment was defined whenever a vehicle changed its active \texttt{headsign}. This criterion was therefore adopted to partition the continuous telemetry streams into individual trip segments for the subsequent predictive modeling analysis.

However, the selected segmentation criterion, as well as the other evaluated features, produced several extreme trip-duration values. These included segments with a duration of 0~min, corresponding to a single observation in the dataset, as well as segments exceeding 20,000~min (more than 13 days), which are clearly inconsistent with the expected duration of a bus trip. Therefore, an additional filtering procedure was applied to restrict the analysis to plausible trip durations.

For the upper bound, an Interquartile Range (IQR)-based filter was applied to the overall distribution of trip durations, resulting in an upper limit of 107.24~min. Applying the same procedure to determine the lower bound yielded a value below 0~min and therefore did not remove any observations, since trip duration cannot be negative. To establish a meaningful lower bound, the IQR criterion was instead evaluated separately for each \texttt{headsign}. The lowest positive lower bound obtained across the individual \texttt{headsign} distributions was 3.5~min, which was adopted as the minimum valid trip duration.

After applying these bounds, the resulting dataset comprised 66,298 valid trip segments, with an average duration of 46.9~min and a median duration of 44.8~min. The relatively close mean and median values indicate that the filtering procedure substantially reduced the influence of the extreme durations observed in the initial segmentation.

With the trip segments established, feature engineering was performed for the predictive framework. Temporal attributes, including the trip start hour, start minute, and weekday (0--6), were incorporated as predictors. The vehicle's direction and headsign were also included as categorical features. The headsign variable was numerically encoded by assigning a unique integer identifier to each distinct destination. Finally, the geodesic distance between the geographic coordinates of the trip's starting and ending points was included as a spatial feature. This measure represents the shortest distance between two coordinates along the Earth's surface, independently of the road network and the actual route followed by the vehicle.

The resulting feature matrix $X$ was divided into training and test sets using an 80:20 split. The training set was used to fit and the test set to benchmark three regression models: Linear Regression, Random Forest Regressor, and eXtreme Gradient Boosting (XGBoost).

Additional features were also evaluated, including vehicle \texttt{id}, \texttt{tripId}, \texttt{lineId}, six-hour periods of the day, weekend or holiday indicator, and the circular standard deviation of the vehicle's heading. However, these variables did not produce consistent improvements in predictive performance and were therefore excluded from the final feature set.

\section{Results and Discussion}

\subsection{Spatiotemporal Demand Profiling and Geometry Mapping}

\begin{figure}[htbp]
\centering
\includegraphics[width=\linewidth]{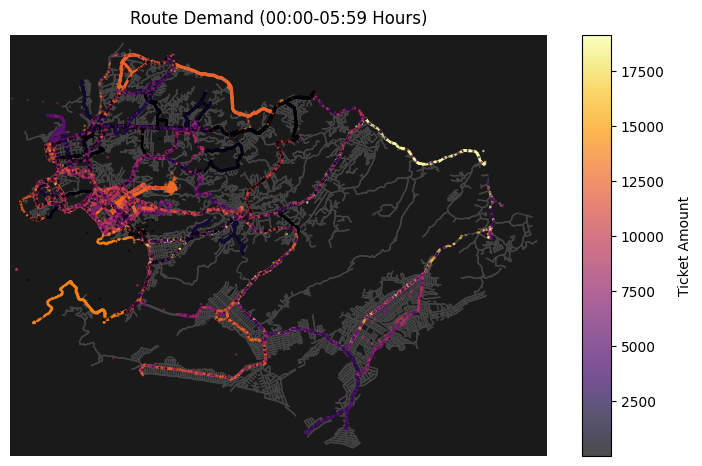}
\caption{Spatial distribution of route-level passenger demand during the 00:00--05:59 operational period.}
\label{fig:21-02}
\end{figure}

\paragraph{Route Demand: 00:00--05:59}
The period from 00:00 to 05:59 exhibits the lowest overall demand in the network, with the maximum values on the map reaching approximately 17,500 ticket transactions per route. Demand is primarily concentrated in the central and western/northwestern areas of the municipality, where a limited number of routes exhibit substantially higher volumes than the remainder of the network. In contrast, routes located farther from the central area generally show lower demand, represented by darker shades of the color scale. This spatial distribution indicates a generalized reduction in public transport usage during the nighttime period, while demand remains concentrated along a smaller number of high-demand corridors. Considering the total number of ticket transactions by hour, demand remains comparatively low from 00:00 to 04:00, before increasing substantially and reaching approximately 190,000 transactions at 05:00.

\begin{figure}[htbp]
\centering
\includegraphics[width=\linewidth]{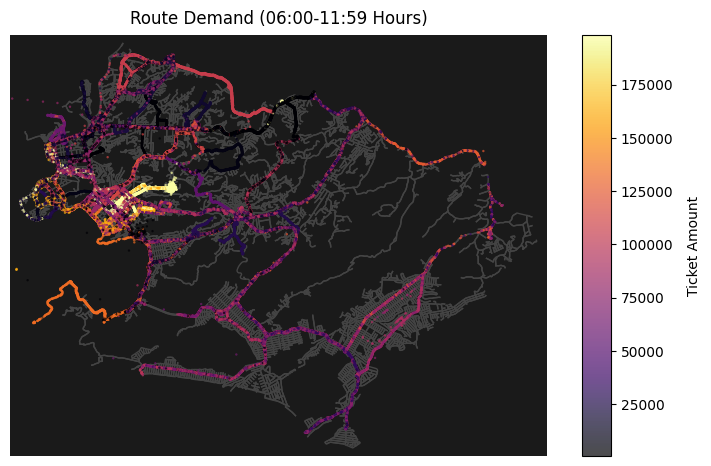}
\caption{Spatial distribution of route-level passenger demand during the 06:00--11:59 operational period.}
\label{fig:03-08}
\end{figure}

\paragraph{Route Demand: 06:00--11:59}
A pronounced increase in demand is observed between 06:00 and 11:59, with the route-level scale exceeding 175,000 ticket transactions. The highest intensities remain concentrated in the central and western/northwestern areas, where several corridors exhibit warmer colors. Compared with the 00:00--05:59 period, the maximum route-level demand is approximately an order of magnitude higher, highlighting the substantial temporal variation in network usage. Higher-demand corridors also extend farther from the central area, indicating a broader spatial distribution of passenger activity during the early morning.

The hourly distribution provides further evidence of this transition. Demand increases sharply from 06:00, reaching approximately 520,000 ticket transactions at 06:00 and 615,000 at 07:00, the highest value observed in the analyzed period. Following this peak, demand gradually decreases, reaching approximately 320,000 transactions at 10:00, before increasing again. Thus, this period is characterized not only by substantially higher demand but also by the expansion of activity across a greater number of network corridors.

\begin{figure}[htbp]
\centering
\includegraphics[width=\linewidth]{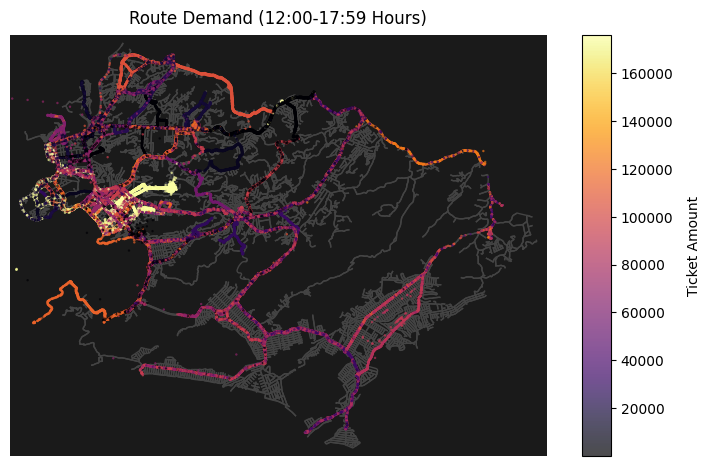}
\caption{Spatial distribution of route-level passenger demand during the 12:00--17:59 operational period.}
\label{fig:09-14}
\end{figure}

\paragraph{Route Demand: 12:00--17:59}
The 12:00--17:59 period maintains one of the highest levels of network demand, with the route-level scale exceeding 160,000 ticket transactions. The highest concentrations remain predominantly in the central and western areas, represented by the lighter shades of the scale. Nevertheless, high-demand corridors are also observed in other parts of the municipality, indicating that substantial passenger activity extends beyond the central area.

The persistence of elevated demand during this period indicates that public transport usage remains high after the early-morning peak. The hourly distribution reinforces this pattern: following a renewed increase beginning around 11:00, demand reaches approximately 475,000 transactions at 12:00, decreases until 14:00, and subsequently rises again to a second major peak of approximately 580,000 transactions at 17:00. Therefore, the 16:00--18:59 interval represents a sustained high-demand period, culminating in the highest afternoon demand observed in the dataset.

\begin{figure}[htbp]
\centering
\includegraphics[width=\linewidth]{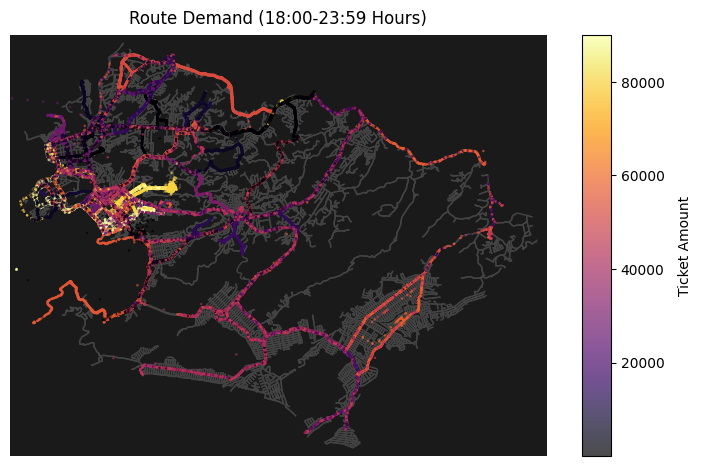}
\caption{Spatial distribution of route-level passenger demand during the 18:00--23:59 operational period.}
\label{fig:15-20}
\end{figure}

\paragraph{Route Demand: 18:00--23:59}
Between 18:00 and 23:59, the route-level demand scale reaches approximately 80,000 ticket transactions, substantially below the maximum values observed during the preceding periods. The central and western areas continue to exhibit the highest concentrations; however, several high-demand corridors are also evident in the southeastern region, represented by warmer colors. This spatial pattern suggests a redistribution of passenger activity across the network during the afternoon and early evening, with specific corridors becoming relatively more prominent.

The hourly distribution shows a pronounced decline following the 17:00 peak. Demand decreases to approximately 490,000 transactions at 18:00, 340,000 at 19:00, and 185,000 at 20:00. From 21:00 onward, the decline becomes even more pronounced, reaching fewer than 40,000 transactions at 23:00 and approximately 15,000 at 00:00. This sharp reduction is consistent with the lower route-level intensities observed in the spatial representation for this period.

\paragraph{Overall Demand Pattern}
Taken together, the maps and hourly distributions demonstrate that demand is highly heterogeneous across both space and time. Rather than being uniformly distributed throughout the public transport network, ticket transactions remain persistently concentrated in the central and western/northwestern areas of Niterói, while specific corridors become comparatively more prominent during particular periods. Temporally, the network exhibits substantial fluctuations in demand, with the 06:00--08:00 and 16:00--18:00 intervals accounting for the highest levels of activity. The results therefore indicate that both the magnitude and spatial distribution of route demand vary considerably throughout the day, with a smaller set of corridors dominating during periods of lower overall demand and a broader portion of the network becoming active during high-demand periods.

\subsection{Operational Stop Validation and Spatial Anomalies}

The analysis indicated substantial spatial and temporal heterogeneity in vehicle dwell events. Most detected events are concentrated along the main transit corridors and in central and densely served areas, with typical dwell times generally ranging from 30 to 120 seconds. In contrast, events exceeding 600 seconds are more spatially concentrated, with notable clusters in the western and northwestern portions of the network and along several major peripheral corridors. These prolonged dwell events suggest locations where vehicles experience recurrent interruptions or extended delays, potentially associated with traffic congestion, intersections, or other operational constraints. Overall, the spatial distribution indicates that dwell behavior is not uniform across the network and may be closely related to variations in demand and local traffic conditions.

Furthermore, a strong spatial correspondence with OSM infrastructure was also observed, with 64.28\% of dwell events occurring within 50~m of at least one OSM-tagged feature. The duration distributions, however, show that the association with different infrastructure types becomes more pronounced among longer dwell events. At the median, dwell durations are highly consistent across most infrastructure categories, ranging from 45 to 46~s, indicating that the most common dwell events exhibit similar durations regardless of the nearby OSM feature. This is consistent with the overall distribution, in which approximately 48\% of all dwell events are concentrated around 45~s.

Greater differences emerge from the 75th percentile onward. While traffic signals, bus stops, and bus stations present $p_{75}$ values of approximately 75--76~s, signalized crossings and crossings reach 90~s, and roundabouts reach 165~s. The divergence becomes even more pronounced in the upper tail: at the $p_{95}$, dwell durations reach 285~s near bus stops, compared with 432~s near traffic signals, 465~s near crossings, 495~s near signalized crossings, and 496~s near roundabouts. These results suggest that OSM infrastructure types are more informative for characterizing prolonged dwell events than typical dwell durations. In particular, the higher upper-percentile values associated with crossings, signalized crossings, and roundabouts may indicate that these locations are more frequently associated with extended vehicle delays.

\begin{figure}[htbp]
\centering
\includegraphics[width=\linewidth]{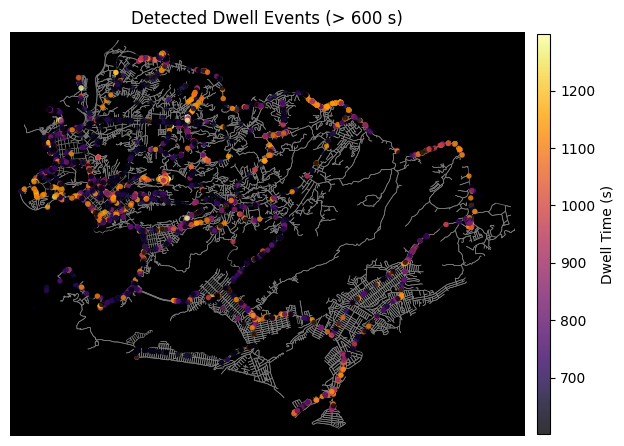}
\caption{Spatial distribution of prolonged vehicle dwell events exceeding 600 seconds across Niterói's bus network.}
\label{fig:dwell_events_600s}
\end{figure}

Additionally, 99\% of the OpenStreetMap points had a nearby dwell event. Also, the number of dwell events associated with each OSM element varied considerably across infrastructure types. Public transport stations had the highest mean number of associated dwell events, with 2,958.59 events per mapped element, followed by bus stations (2,085.70), signalized crossings (1,919.79), and bus stops (1,649.48). Crossings and traffic signals presented mean values of 1,462.59 and 1,188.60 associated events, respectively, while roundabouts had the lowest mean, with 618.77 events per element. These differences reflect the varying prevalence and spatial distribution of each infrastructure type throughout the analyzed network.

Finally, a clear temporal pattern was observed across the week. Weekdays recorded approximately 60,000--75,000 dwell events per day, nearly twice the 30,000--35,000 observed on weekends. The number of events increased from Sunday through Wednesday, remained relatively stable from Wednesday to Friday, and then declined toward the weekend. This temporal variation is consistent with the greater operational activity and demand typically observed on weekdays. Overall, the results demonstrate that the detected dwell events capture meaningful temporal differences in bus operations, complementing the spatial patterns identified across the network.

\subsection{Continuous Trip Segmentation and Predictive Modeling}

The supervised learning framework demonstrated strong performance in predicting transit trip durations. The baseline Linear Regression model achieved a Mean Absolute Error (MAE) of 11.65 minutes, a Root Mean Squared Error (RMSE) of 14.71 minutes, and a coefficient of determination ($R^2$) of 0.45. The ensemble-based models substantially improved predictive performance, indicating their ability to capture the non-linear relationships and complex spatiotemporal interactions underlying transit travel times. The Random Forest Regressor reduced the MAE to 5.23 minutes and the RMSE to 8.06 minutes, while achieving an $R^2$ of 0.835. XGBoost yielded the best overall results, with an MAE of 5.22 minutes, an RMSE of 7.90 minutes, and an $R^2$ of 0.84.

The contribution of each feature to trip-duration prediction was further assessed through an ablation analysis, in which features were removed from the XGBoost model while keeping the remaining configuration unchanged. Removing \texttt{direction} resulted in only a modest degradation, increasing the MAE to 5.50 minutes and reducing $R^2$ to 0.828, suggesting that direction provides useful but relatively limited additional information. In contrast, removing \texttt{weekday} increased the MAE to 6.67 minutes and reduced $R^2$ to 0.768, indicating a substantially greater contribution to predictive accuracy. The removal of \texttt{distance\_m} produced a further decline, with MAE increasing to 6.92 minutes, RMSE to 10.43 minutes, and $R^2$ decreasing to 0.724. Temporal information was also highly influential: removing both \texttt{start\_hour} and \texttt{start\_minute} increased the MAE to 7.38 minutes and reduced $R^2$ to 0.724. Finally, removing \texttt{headsign} produced the largest deterioration in performance, with MAE increasing to 7.77 minutes, RMSE to 10.80 minutes, and $R^2$ decreasing to 0.704. Overall, the ablation results indicate that \texttt{headsign}, temporal attributes, \texttt{distance\_m}, and \texttt{weekday} provide the most substantial contributions to trip-duration prediction, whereas \texttt{direction} has the smallest marginal contribution among the evaluated features.

Overall, these results indicate that integrating route-related characteristics with temporal features can substantially improve the prediction of transit trip durations, providing a promising foundation for applications such as real-time fleet management, operational planning, and passenger travel-time estimation.

Additionally, model performance was evaluated separately for each headsign to assess whether trip duration could be predicted consistently across individual destinations. The same feature set used in the previous predictive analysis was retained, while XGBoost was selected as the regression model based on its superior performance in the overall evaluation. To ensure a sufficient number of observations for model training, only headsigns associated with more than 500 trips were considered, resulting in 39 individually trained models. Across these models, the mean MAE was 5.68 minutes (SD = 1.23), while the mean RMSE was 8.66 minutes (SD = 1.58). The mean coefficient of determination was $R^2 = 0.61$ (SD = 0.16), indicating moderate predictive performance overall.

However, substantial variation was observed across destinations. The MAE ranged from 3.15 to 8.67 minutes, with a median of 5.80 minutes, while the RMSE ranged from 5.56 to 12.51 minutes, with a median of 8.56 minutes. The $R^2$ values ranged from 0.251 to 0.829, with a median of 0.648. Thus, although the aggregated model achieved an $R^2$ of 0.84, the headsign-specific models exhibited greater variability in predictive performance. This difference suggests that trip-duration predictability is not uniform across destinations and may depend on route-specific operational and traffic characteristics.

\section{Conclusion and Future Work}

This work established a robust data fusion and predictive framework by synchronizing heterogeneous passenger boarding transactions with high-frequency GPS telemetry across Niterói’s bus network. Our spatial analyses successfully mapped highly monocentric demand structures (Objective 1) and isolated critical roadway friction points by parameterizing structural dwell events (Objective 2).

Addressing the core predictive task (Objective 3), we developed a multi-criteria trip segmentation algorithm to overcome the temporal inconsistencies associated with the native itinerary indices. By integrating these engineered temporal, directional, and route-specific features into supervised learning models, we achieved highly precise travel duration forecasts.

Future developments may focus on deploying these trained architectures within dynamic, real-time prediction pipelines to feed passenger-facing arrival systems. Additionally, we aim to incorporate external environmental streams, such as real-time weather logs and sudden road event notifications, to enhance model robustness against unpredicted roadway disruptions.

\bibliography{sample}

\end{document}